\documentclass[
reprint,
superscriptaddress,
 amsmath,amssymb,
 aps,
 pra,
 showkeys,
floatfix,
]{revtex4-2}
\usepackage[utf8]{inputenc}
\usepackage{graphicx}
\usepackage{dcolumn}
\usepackage{bm}
\usepackage{bbold}
\usepackage{physics}
\usepackage[colorlinks=true, allcolors=blue]{hyperref}
\usepackage{siunitx}
\usepackage{mathdots}
\usepackage[dvipsnames]{xcolor}
\usepackage{comment}
\DeclareSIUnit\dBm{dBm}

\newcommand{\matr}[1]{\bm{#1}}

\newcommand{\cov}[1]{\text{cov}\left[{#1}\right]}

\definecolor{drkgreen}{rgb}{0.0, 0.5, 0.0}
\definecolor{violet}{rgb}{0.5, 0.0, 1.0}

\begin{document}
\preprint{APS/123-QED}
\title{Continuous-variable two-dimensional cluster states in the microwave domain}%

\author{Fabio Lingua}
 \email{lingua@kth.se}
 \affiliation{Department of Applied Physics, KTH Royal Institute of Technology, SE-10691 Stockholm, Sweden}

\author{Michele Cortinovis}
 \affiliation{Department of Applied Physics, KTH Royal Institute of Technology, SE-10691 Stockholm, Sweden}
 \affiliation{Dipartimento di Fisica, Politecnico di Milano, I-20133 Milano, Italy}
 
\author{J. C. Rivera Hernández}
 \affiliation{Department of Applied Physics, KTH Royal Institute of Technology, SE-10691 Stockholm, Sweden}

\author{David B. Haviland}
 \affiliation{Department of Applied Physics, KTH Royal Institute of Technology, SE-10691 Stockholm, Sweden}

\date{\today}

\begin{abstract}
We demonstrate the experimental realization of two-dimensional, continuous variable (CV) cluster states between 191 microwave frequency modes. 
This result is obtained by exposing vacuum fluctuations to the input of a Josephson Parametric Amplifier, parametrically pumped by a sum of coherent tones around twice its resonant frequency. 
By carefully tuning pump frequencies, amplitudes, and phases we engineer the interference between mixing products and realize honeycomb and square lattice CV cluster states with three and four pump tones respectively. 
We prove the presence of the cluster states with a suitable nullifier test, reaching up to $-1.2$~dB of squeezing of the cluster state's nullifiers. 
We study hidden entanglement (HE) and show no hidden entanglement up to $\sim -1$~dB of squeezing and negligible HE at optimal squeezing.
\end{abstract}


\maketitle

\section{Introduction}
\begin{figure}[h!]
    \centering
    \includegraphics[width=\columnwidth]{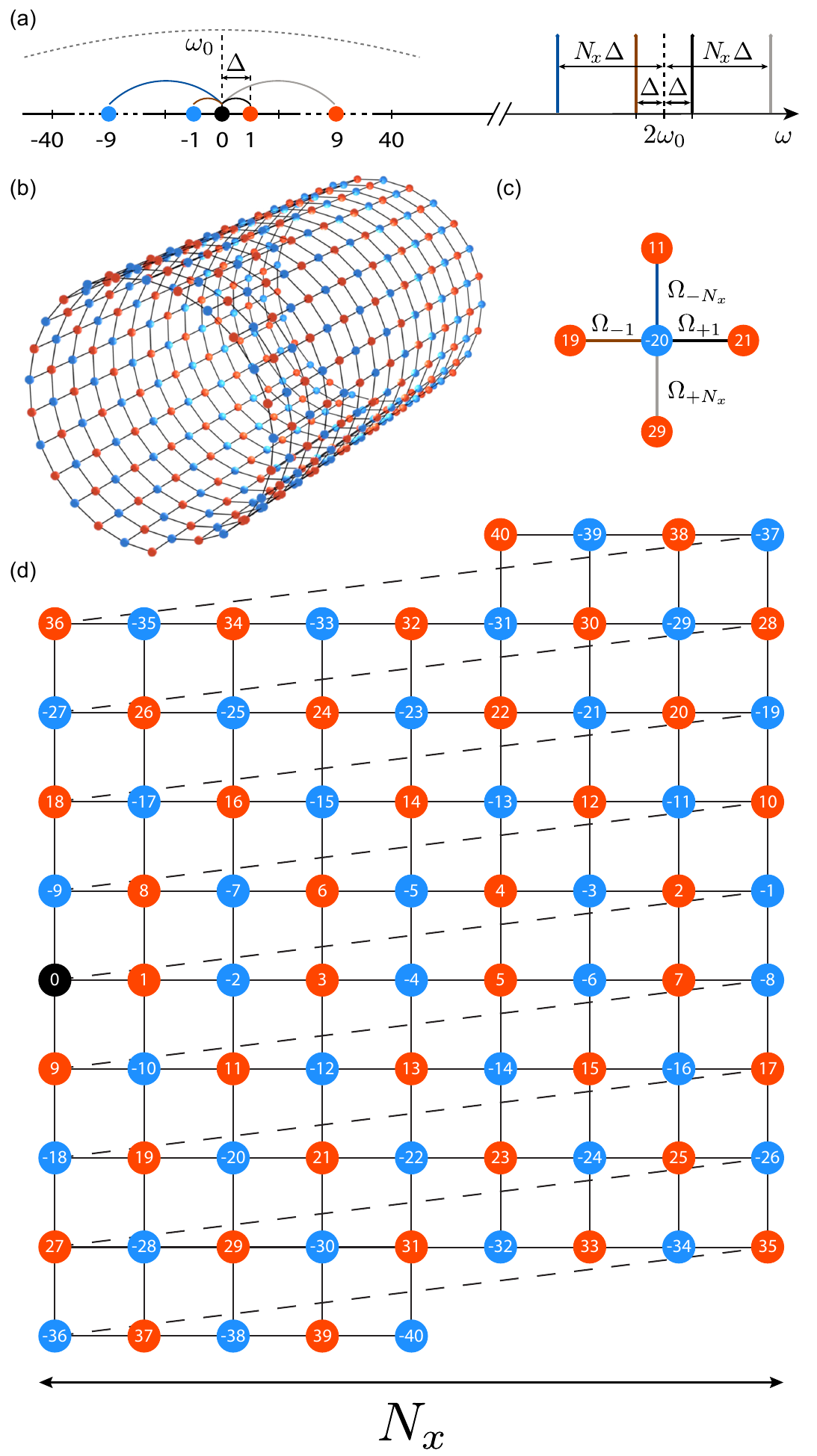}
    \caption{(a) Equally spaced modes (spacing $\Delta$) around $\omega_0$ and pumps near $2\omega_0$ select pairwise couplings.
    (b) Resulting square-lattice graph with periodic boundary (cylinder).
    (c) Nearest-neighbour connectivity of mode $\Omega$ to $\Omega_{\pm1}$ and $\Omega_{\pm N_x}$.
    (d) Unwrapped lattice with periodic links.}
    \label{fig:CVCS_Sq}    
\end{figure}

Continuous-variable measurement-based quantum computing (CV-MBQC) has emerged as a promising paradigm for scalable quantum information processing~\cite{briegel_measurement-based_2009, menicucci_universal_2006}. 
In this approach, quantum computation is performed by local measurements on a highly entangled multipartite resource state known as a cluster state~\cite{raussendorf_one-way_2001, raussendorf_measurement-based_2003}. 
These states are characterized by correlations between the quadratures of multiple bosonic modes and they can be naturally represented as a \emph{canonical graph}~\cite{menicucci_universal_2006, menicucci_one-way_2008, gu_quantum_2009, zhang_continuous-variable_2006, weedbrook_gaussian_2012}. 
The ability to engineer and control the connectivity of such graphs constitutes a central requirement for the practical implementation of CV-MBQC. 
Theoretical studies have established that a two-dimensional square-lattice graph as shown in Fig.~\ref{fig:CVCS_Sq}, provides a universal resource for quantum computation~\cite{menicucci_universal_2006, menicucci_fault-tolerant_2014}.

Continuous-variable cluster states of medium and large scale have been successfully demonstrated at optical frequencies~\cite{menicucci_one-way_2008, chen_experimental_2014, pfister_continuous-variable_2020}, using time-multiplexing techniques~\cite{yukawa_experimental_2008, menicucci_arbitrarily_2010, menicucci_temporal-mode_2011, yokoyama_ultra-large-scale_2013, cai_multimode_2017, asavanant_generation_2019, larsen_deterministic_2019, madsen_quantum_2022, yokoyama_full-stack_2025}.
By contrast, the cluster state generation at microwave frequencies remained considerably more challenging.
While the techniques used at optical frequencies are difficult to apply and scale in the microwave domain, digital signal processing offers an alternative and more viable path forward.
Experimental efforts in the microwave domain have demonstrated multimode Gaussian entanglement and cluster-like correlations~\cite{menzel_path_2012, jolin_multipartite_2023, hernandez_control_2024}, but have so far been restricted to systems involving a discrete and relatively small number of modes~\cite{eichler_observation_2011,petrovnin_generation_2023, alocco_programmable_2025}. 

The first large-scale experimental realization in the microwave domain is the generation of square-ladder cluster states of $94$ frequency modes~\cite{lingua_continuous-variable_2025}, and with discrete variables between $16$ transmon qubits~\cite{osullivan_deterministic_2025}.
These works represent an important intermediate step but still lack the full two-dimensional connectivity required for universal MBQC. 
Extending these architectures toward genuine two-dimensional cluster states poses a conceptual challenge, as it requires a systematic understanding of how a desired graph adjacency translates into physically realizable correlations and, consequently, into a suitable covariance matrix~\cite{menicucci_graphical_2011}. 
Addressing this mapping is a prerequisite for the controlled design of two-dimensional cluster states~\cite{van_loock_building_2007}.

Two important figures of merit describe the quality of a cluster state:  the strength of quantum correlations as measured by the variance of the nullifier squeezing below its vacuum level, and some measure of the hidden entanglement revealing correlations beyond the ideal graph structure~\cite{ohliger_limitations_2010, gonzalez-arciniegas_cluster_2021}. 
Such correlations do not define the target cluster-state connectivity but can nevertheless affect the structure and quality of the resource state. 

In this work, we develop a framework for engineering and characterizing large-scale microwave continuous-variable cluster states with two-dimensional connectivity. 
Our approach enables the design of nontrivial graph topologies, including square and honeycomb lattices. 
These results establish a pathway toward scalable continuous-variable cluster states in superconducting quantum systems, bridging the gap between optical and microwave implementations of measurement-based quantum computation.

\section{CV Cluster states}

Continuous-variable quantum information uses \emph{qumodes}, eigenstates of the quadrature operators $x,p$, instead of qubits as fundamental unit of information.
Continuous-variable cluster states (CVCS) are multipartite entangled states defined on a \emph{canonical graph} whose nodes are represented by qumodes.  Links between nodes (graph edges) are established through the CV-equivalent of the controlled-Z gate, or operators  $e^{i A_{ij}  x_i x_j}$, which displace the momentum of mode $j$, conditioned on the value of the position of mode $i$.

A CVCS between $N$ qumodes is formally defined as
\begin{equation}
    \ket{\Psi}=C_\text{z} \ket{0}_{p_1}\ket{0}_{p_2}\dots\ket{0}_{p_N}, \;\;\;\label{CVCS}
\end{equation}
where the qumodes are initialized in the equal, continuous superposition described by the zero-momentum eigenstates $\ket{0}_{p_i}=\int dx_i \ket{x_i}$. 
The operator $C_z = \prod_{i,j}^Ne^{i A_{ij}  x_i x_j}=e^{i\vec{x}^T\matr{A}\vec{x}}$ combines the controlled-z gate on all qumodes pairs.
The terms $A_{ij}$ are elements of the adjacency matrix $\matr{A}$ that describes the topology of the canonical graph, defining the pairwise controlled-z link between connected modes~\cite{menicucci_graphical_2011, pfister_continuous-variable_2020}.
A CVCS on $N$ qumodes is defined as a stabilizer state, an eigenstate of the $N$ generators of the stabilizer group with eigenvalue 1. The generators are defined $\mathcal{S}_i=e^{iN_i}$, with $N_i=p_i- \sum_jA_{ij}\,x_j$, with the nullifier operators satisfying
\begin{eqnarray}
    N_i\ket{\Psi} &=& 0\ket{\Psi},\\ 
    \Delta N_i^2 &=& 0,
    \label{null}
\end{eqnarray}
which implies  perfect correlation or infinite squeezing between the quadratures of modes that are connected on the canonical graph.
Cluster states as defined through~\eqref{CVCS} cannot be realized exactly in an experiment.
Realization of CVCS is possible at finite squeezing using a close approximation to quadrature eigenstates, Gaussian squeezed-vacuum states.
For these states, the adjacency matrix is replaced by the complex matrix
\begin{equation}
    \matr{Z} = \matr{A} + i\matr{U},
    \label{eq:Z}
\end{equation}
where $\matr{U}$ is a diagonal matrix that contains the finite squeezing contribution to the state. It describes a Gaussian envelope of the distribution of the $N$ quadratures $\vec{x}=(x_1,x_2,\dots,x_N)^T$.
\begin{equation}
    \ket{\Psi}= e^{-\vec{x}^T\matr{U}\vec{x}}e^{i\vec{x}^T\matr{A}\vec{x}}\ket{0}_{p_1}\ket{0}_{p_2}\dots\ket{0}_{p_N}, \;\;\; \label{gauss_CVCS}
\end{equation}
The joint stabilizer condition for finite-squeezed cluster states transforms to
\begin{equation}
    \left(\vec{p} - \matr{A}\vec{x}\right)\ket{\Psi}=i\matr{U}\vec{x}\ket{\Psi}
\end{equation}
which is equivalent to the stabilizing condition~\eqref{null} in the limit $\matr{U}\rightarrow\mathbb{0}$, i.e. infinite squeezing. 
For finite squeezing $\Delta N_i^2\leq \Delta N_{i\,0}^2$, where $\Delta N_{i\,0}^2=\bra{0}(N_i - \langle N_i\rangle)^2\ket{0}$ is the nullifier variance for the vacuum state. 

Gaussian states are completely described by the first and second statistical moments of quadratures. 
The second moments are collected in the covariance matrix,
\begin{equation}
    \cov{\vec{q}}_{ij}=\frac{1}{2}\bra{\Psi}\{q_i,q_j\}\ket{\Psi},    
\end{equation}
where $\{\cdot,\cdot\}$ denotes the anti-commutator between quadrature operator components $i$ and $j$ of a quadrature vector $\vec{q}$. 

Let us denote the covariance matrix in the mode ordering $\vec{q}=(x_1,p_1,x_2,p_2,\dots,x_N,p_N)^T$ as $\matr{V}$, and the covariance matrix in the quadrature ordering $\vec{r} = (\vec{x}, \vec{p})^T$ as $\matr{V_r}$. 
The statistical properties of the Gaussian cluster state~\eqref{gauss_CVCS} are naturally encoded in either representation of the covariance matrix. 
Using a Wigner-Moyal-like decomposition~\cite{simon_gaussian_1988,menicucci_graphical_2011} to represent $\matr{V_r}$,  
\begin{equation}
    \matr{V_r}=\frac{1}{2}\begin{pmatrix}
                    \matr{U}^{-1} & \matr{U}^{-1}\matr{A}\\
                    \matr{A}\matr{U}^{-1} & \matr{U} + \matr{A}\matr{U}^{-1}\matr{A}
                \end{pmatrix}
        \label{Eq:Covariance_graphical}
\end{equation}
the matrices $\matr{A}$ and $\matr{U}$ can be easily computed.
In this work, we use a normalized definition of the adiecency matrix $\matr{A}$ where $A_{ij}=\{0,\pm 1\}$ to compute the canonical graphs.
The first quadrant of $\matr{V_r}$ corresponds to $\tfrac{1}{2}\matr{U}^{-1}=\cov{\vec{x}}$ which contains the correlations between the $\vec{x}$ quadratures. 
In the limit of very strong squeezing  $\matr{U}\rightarrow \mathbb{0}$ (i.e. $\text{Tr}[\matr{U}]\rightarrow 0$), non-zero element of $\matr{U}^{-1}$ may still result in very strong \emph{hidden} correlations~\cite{gonzalez-arciniegas_cluster_2021}.
These hidden correlations, or leakage of entanglement towards unwanted modes, seriously compromise the quality of the final cluster state.
An additional more stringent condition on $\matr{U}$ to prevent leakage of entanglement follows from the definition of Gaussian cluster states~\eqref{gauss_CVCS} where the matrix $\matr{U}$ is required to be dominant diagonal (implying the diagonality of $\matr{U}^{-1}$).
\begin{figure*}[t]
    \centering
    \includegraphics[width=\textwidth]{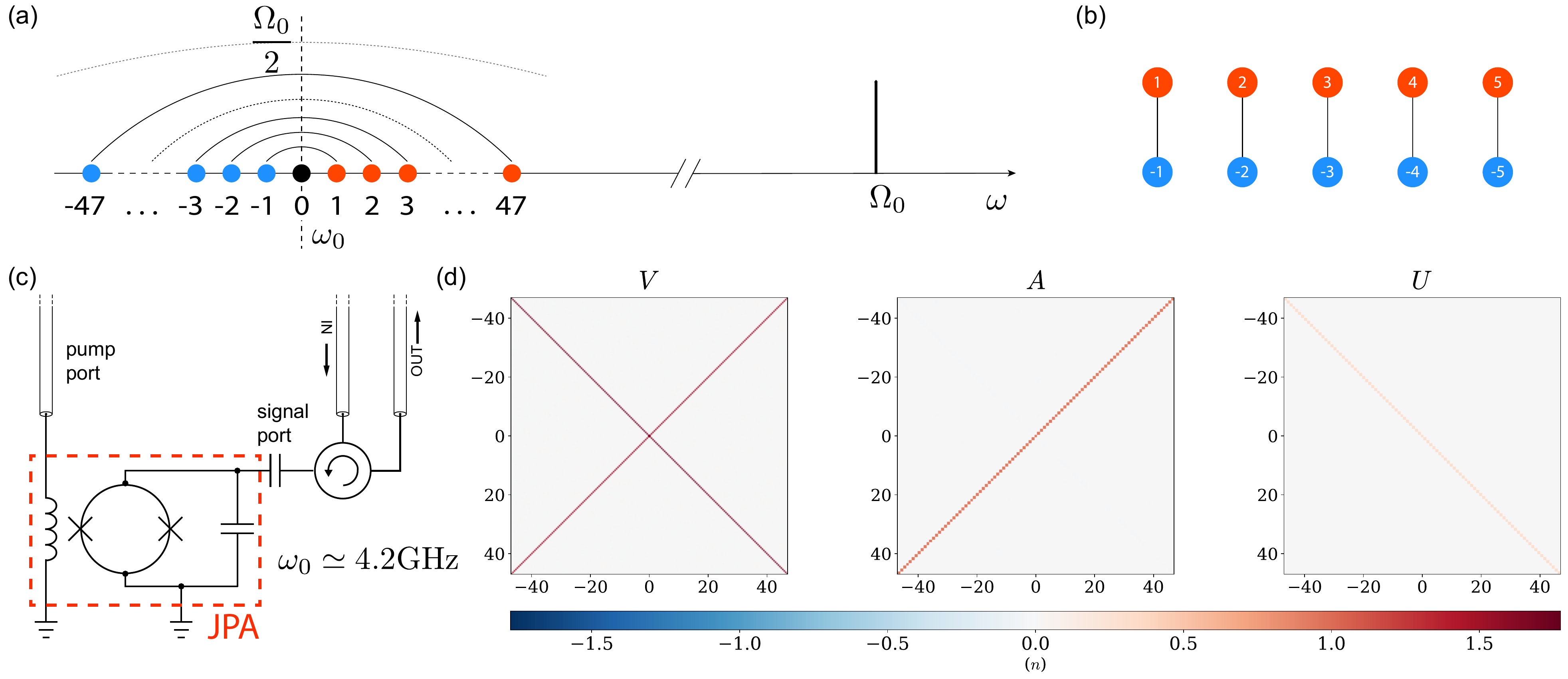}
    \caption{
    (a) Frequency-comb architecture centered at $\omega_0$. The parametric pump at $\Omega_0 \simeq 2\omega_0$ mediates correlations between opposite modes. 
    (b) Mode pairing structure illustrating two-mode squeezing interactions $(i,-i)$. 
    (c) Schematic of the Josephson Parametric Amplifier (JPA) with separated pump and signal ports; the device is operated around $\omega_0 \simeq 4.2$\,GHz. 
    (d) Representative covariance matrix $\matr{V}$, adjacency matrix $\matr{A}$, and $\matr{U}$ matrix (in photon-number units), highlighting the characteristic cross-diagonal structure of the engineered cluster-state correlations.
    }
    \label{fig:VAU_1pmp}    
\end{figure*}

\section{Microwave cluster state generation}

A Josephson Parametric Amplifier (JPA) is a microwave parametric oscillator whose resonance frequency is tunable by an external pump.  
The pump consists of a DC bias which fixes the static resonance to $\omega_0\simeq4.2$~GHz, and an RF waveform with multiple discrete frequency components, all close to $2\omega_0$. 
The JPA is capacitively coupled to a circulator which separates the input and output propagating modes (see Fig.~\ref{fig:VAU_1pmp}c) for a schematic diagram). 
Vacuum fluctuations are injected at the input, and the output is amplified with a cryogenic low-noise amplifier and digitally sampled.  
The pump is synthesized by a multifrequency lock-in amplifier which also demodulates the output, giving both quadratures of the response at $N$ orthogonal frequencies $\omega_i$ equally spaced around $\omega_0$ by $\Delta=\frac{2\pi}{T}$, where $T$ is the measurement time window (see appendix~\ref{ExpSetup} for detailed description of the experimental setup). 
The $N$ demodulated frequencies constitute the orthogonal basis of qumodes from which we build the cluster state. 

The pump waveform is periodic on the time window $T$, and it is composed of a superposition of coherent tones
\begin{equation}
    g_p(t)=\sum_k g_k\cos(\Omega_kt + \phi_k), \label{eq:pump}
\end{equation}
where $\Omega_k = 2\omega_0 + k\Delta$.
Modulation of the parametric oscillators resonant frequency through~\eqref{eq:pump} generates intermodulation.  Each pump tone mixes frequencies which are symmetric around $\Omega_k / 2$. 
With multiple pump frequencies, interference between different mixing processes connecting the same two modes, enables the design of correlations between their vacuum fluctuations. 
We structure these correlations by adjusting amplitudes $g_k$ and phases $\phi_k$ at tuned pump frequencies $\Omega_k$, thereby synthesizing the canonical graph and adjacency matrix of the CV cluster state~\eqref{gauss_CVCS}.

In our previous work~\cite{lingua_continuous-variable_2025} we demonstrated the experimental realization of a CV cluster state (CVCS) between $N=94$ microwave frequency modes in a canonical graph with square-ladder topology, having a nullifier with $-1.4$~dB of squeezing below vacuum. 
This CVCS was realized with parametric pumps at three frequencies, each having a specific phase relation. 
The relative phase of the pumps plays a crucial role, enabling destructive interference between different mixing processes having the same order of intermodulation, effectively canceling unwanted correlations.

The square ladder is not fully two-dimensional as increasing the number of modes causes the lattice to grow in only one dimension (i.e., length of the ladder).
Engineering a fully two-dimensional canonical graph requires mapping the one-dimensional frequency axis of discrete modes onto a two-dimensional graph.
Arbitrary permutation of the modes between the nodes of the graph may result in a mapping that is not realizable through parametric pumping. 
The challenge is understanding what mappings are possible.
\begin{figure*}[t]
    \centering
    \includegraphics[width=\textwidth]{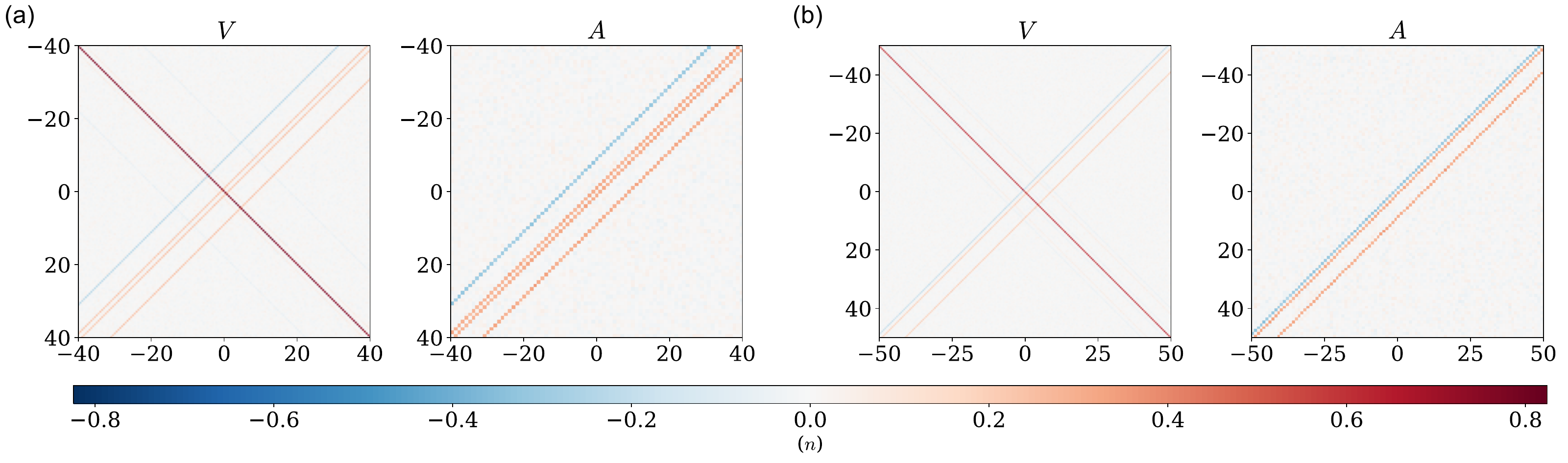}
    \caption{
    (a) Square-lattice configuration: measured covariance matrix $\matr{V}$ and corresponding adjacency matrix $\matr{A}$. The structure of $\matr{A}$ reflects the target graph connectivity. 
    (b) Honeycomb-lattice configuration: measured $\matr{V}$ and reconstructed $\matr{A}$, showing the characteristic pattern of the engineered graph. Color scale in photon-number units.
    }
    \label{fig:VandA_SqHC}    
\end{figure*}

\subsection{Two-mode vs. multi-mode correlations}
Consider a single pump at frequency $\Omega \simeq 2 \omega_0$.  We label the modes as integers, with zero at half the pump frequency $\tfrac{\Omega_0}{2}$, negative modes blue, and positive modes red. A single pump connects positive modes to negative modes in a pairwise fashion $\omega_i\leftrightarrow\omega_{-i}$. 
The resulting covariance matrix $\matr{V}$ displays a main anti-diagonal as shown in Fig.~\ref{fig:VAU_1pmp}d).
From the covariance matrix we extract both the adjacency matrix $\matr{A}$ and the diagonal matrix $\matr{U}$  as shown in panel d).
These matrices fully describe $\tfrac{N-1}{2}$ independent two-mode CVCS's with the canonical graphs in Fig.~\ref{fig:VAU_1pmp}b).
These simple two-mode or zero-dimensional graphs do not contain hidden entanglement or unwanted edges connecting modes that we do not want to correlate.

When the pump has multiple frequency components, the graph connectivity becomes much more complex.
Higher-order intermodulation processes, where multiple pump frequencies mix recursively, result in off-diagonal as well as anti-diagonal structure in the covariance matrix. 
In~\cite{hernandez_control_2024,lingua_continuous-variable_2025} we showed that by tuning the relative phase of three pumps it is possible to make higher order mixing products destructively interfere, effectively canceling unwanted correlations. 
In the frequency domain, the parametric mixing process is a convolution of the pump spectrum with each frequency mode~\cite{hernandez_control_2024}, giving an intrinsic translational symmetry to the mode connections.
The result is an intertwined and periodic arrangement of links between modes.
The unknotting of these links to determine the topology of the corresponding canonical graph is, in general, a hard problem to solve. 
   
To realize a fully 2D canonical graph we start from the notion that a parametric pump around $2\omega_0$ always connects positive modes to negative modes. 
Imagine a checkerboard-like arrangement of modes where every positive mode is nearest-neighbor to only negative modes, and vice versa.
By carefully choosing the position of the measurement mode basis and placing the frequency components of the parametric pump around $2\omega_0$, we can engineer the connectivity between positive and negative modes and realize connections along the horizontal and vertical lattice directions. 

\begin{figure}
    \centering
    \includegraphics[width=\columnwidth]{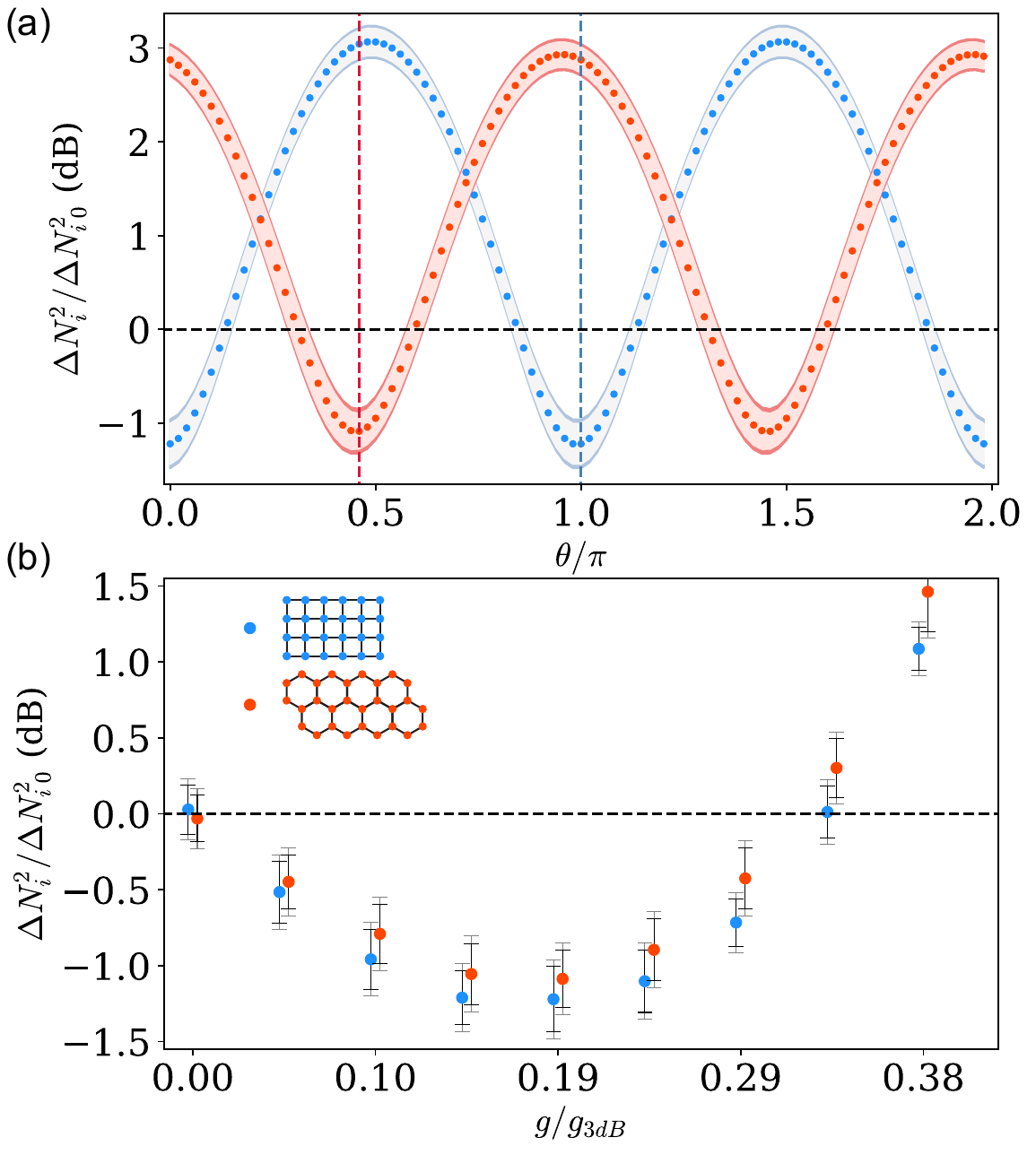}
    \caption{
    (a) Normalized nullifier variance $\Delta N_i^2 / \Delta N_{i,0}^2$ as a function of the measurement angle $\theta$, showing alternating squeezing and anti-squeezing. Dashed lines mark the optimal angles. 
    (b) Minimum nullifier variance versus normalized pump strength $g/g_{3\mathrm{dB}}$ for square (blue) and honeycomb (orange) lattices. Error bars denote experimental uncertainty; the dashed line indicates the vacuum level.
    }
    
    \label{fig:Null_SqHC}
\end{figure}

\subsection{Square lattice}

With four parametric pumps we can realize the canonical graph having a square-lattice topology with $N$ frequency modes and tunable width $N_x$.
Figure~\ref{fig:CVCS_Sq}a) shows the pumping scheme that realizes the single-mode graph shown in Fig.~\ref{fig:CVCS_Sq}c), generating the square lattice canonical graph shown in Fig.~\ref{fig:CVCS_Sq}d). 
Four pumps of identical amplitude, $g_k=g$ $\forall k$ are placed at $\Omega_{\pm1}$, realizing the horizontal connections in the graph. 
The generic mode $i$ is connected horizontally to the modes $-i\pm 1$. 
The pumps at frequency $\Omega_{\pm N_x}$ connect the nodes of the graph vertically, expanding the graph to two dimensions. 
The mode $i$ is now connected to modes $-i \pm N_x$. 
The $N_x\Delta$ spacing between pump frequencies determines the horizontal size of the lattice. 

The combined action of all pumps results in a square lattice with periodic boundary conditions, where the right boundary is connected to the next row of the left boundary.  
The resulting lattice is therefore a chiral cylindrical manifold with circumference $N_x$.
Figure~\ref{fig:CVCS_Sq} b) shows the 3D representation of the two-dimensional square-lattice of panel d).
The phases of the four pumps are set to zero, with the exception of the pump at $k=-N_x$, which is set to $\pi$, $\phi_{-Nx}= \pi$.
This phase configuration cancels correlations which would otherwise connect next-nearest neighbors on a diagonal of the square lattice~\cite{hernandez_control_2024,lingua_continuous-variable_2025}. 

\begin{figure*}[t]
    \centering
    \includegraphics[width=\textwidth]{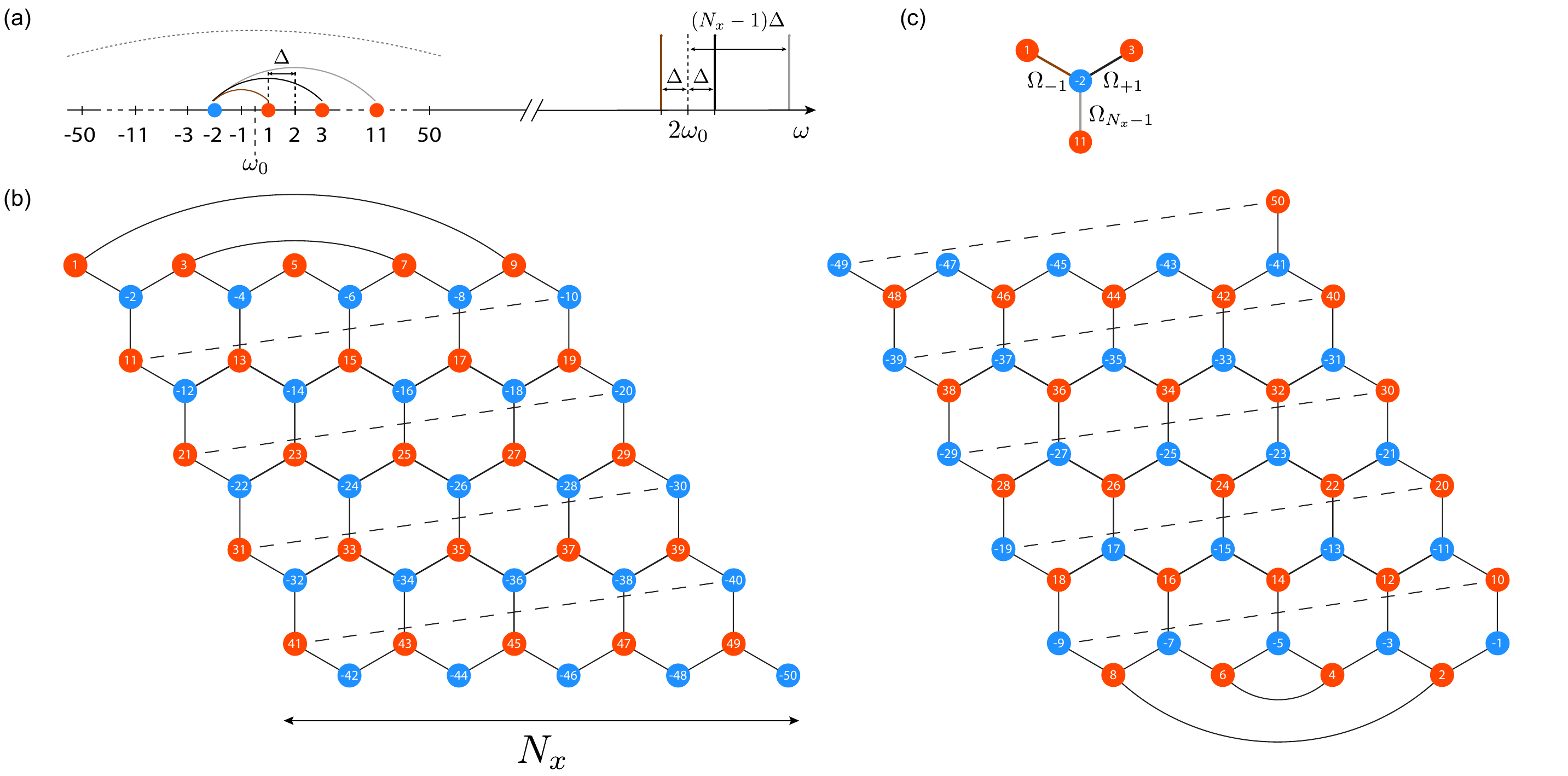}
    \caption{
    (a) Frequency-comb arrangement around $\omega_0$ with spacing $\Delta$ and parametric pumping around $2\omega_0$. 
    (b) Mode mapping to a two-dimensional honeycomb lattice of width $N_x$, obtained from nearest-neighbor couplings between symmetrically detuned modes. 
    (c) Local connectivity pattern generated by the pump tones $\Omega_k$, defining the graph links in frequency space.
    }
    \label{fig:CVCS_HC}    
\end{figure*}

Applying the pumping scheme shown in Fig.~\ref{fig:CVCS_Sq}a) to the JPA with vacuum fluctuations at its input, we generate quantum correlations which we characterize through measurement of the quadratures at each frequency. 
We analyze $10^6$ consecutive time windows with care taken to keep a fixed phase reference for all time windows and all frequencies. 
The covariance matrix $\matr{V}$ is built by averaging products of the quadratures of different frequencies. 
See appendix~\ref{app_CovMeas} and supplemental material~\cite{lingua_supplemental_2026} for more details on the phase reference, noise calibration, and construction of the covariance matrix. 

The measurement is repeated at multiple values of the pump amplitude $g$.
Figure~\ref{fig:VandA_SqHC}a) shows the measured covariance matrix $\matr{V}$ and adjacency matrix $\matr{A}$ for the square lattice cluster state at the pump power and quadratures rotation angle giving the lowest nullifier value, or most squeezing (see blue vertical dashed line in Fig.~\ref{fig:Null_SqHC}a) described below). 
The covariance matrix exhibits four main anti-diagonals confirming correlations between frequencies connected directly by the four pumps (i.e., three-wave mixing processes). 
As we shall see in the analysis of $\matr{U}$, not all the higher-order intermodulation products are canceled. 
Some unwanted correlations are measurable at higher pump powers, although they are significantly weaker than the desired correlations. 

\subsection{Honeycomb lattice}

Following the same approach as the square lattice, a honeycomb lattice graph with $N_x$ horizontal modes is realized with three pumps.  
Three edges are obtained with a combination of two pumps at $\Omega_{\pm1}$ realizing the oblique edges, and a single pump at $\Omega_{N_x-1}$ realizing the vertical edges. 
The checkerboard pattern of positive and negative modes requires the same pump at $\Omega_{N_x-1}$ to connect both positive and negative modes in the vertical direction.
The vertical connection is effectively achieved by altering the mapping to remove mode $0$ on the frequency axis, by placing $\omega_0$ exactly in between two modes in the comb, as shown in Fig.~\ref{fig:CVCS_HC}a).
This mapping requires the reference pump frequency $2 \omega_0$ to be an odd-integer multiple of $\Delta$, as opposed to an even integer for the case of the square lattice.
The result is two disconnected honeycomb lattices shown in Fig.~\ref{fig:CVCS_HC}b), for $N_x=10$ and $N=50$. 
The single-mode graph is shown in Fig.~\ref{fig:CVCS_HC}c).
The pump at $\Omega_{N_x-1}$ controls the width of the lattice and induces a chiral periodic boundary conditions in the horizontal direction. 
This pump also closes or caps one end of both lattices. 
The vertical length of the honeycomb lattices depends on the total number of measured modes $N$. 

Using the same procedure as for the square lattice, we performed the experiment on the JPA with the pumping scheme shown in Fig.~\ref{fig:VandA_SqHC}a).
All three pumps were set to the same strength $g$, and the phase of one of the pumps was set to $\pi$ to cancel higher-order, overlapping intermodulation products.
Figure~\ref{fig:VandA_SqHC} b) shows the measured covariance matrix $\matr{V}$ and adjacency matrix $\matr{A}$ at the pump amplitude which achieves maximum nullifier squeezing $g/g_{3\text{dB}}=0.19$.

\subsection{Nullifier analysis}

A comparison of the nullifier squeezing for both square and honeycomb canonical graphs is shown in Fig.~\ref{fig:Null_SqHC}. 
Figure~\ref{fig:Null_SqHC}a) shows the nullifier variance averaged over all modes $\Delta N^2_i/\Delta N^2_{i\,0}$ as a function of the global rotation angle $\theta$ of all quadratures in the covariance matrix (see appendix~\ref{app_CovMeas} for details). 
As expected, squeezing to anti-squeezing is $\pi$ periodic in the rotation angle of the quadrature basis. 
At the optimal angle $\theta$ we reach $-1.22$~dB of nullifier squeezing for the square lattice and $-1.08$~dB for the honeycomb, or $-4.68$ and $-4.58$ standard deviations below the vacuum level respectively.
These measurements verify the generation of two-dimensional microwave CV Cluster States with both square and honeycomb lattice topologies.

\begin{figure}
    \centering
    \includegraphics[width=\columnwidth]{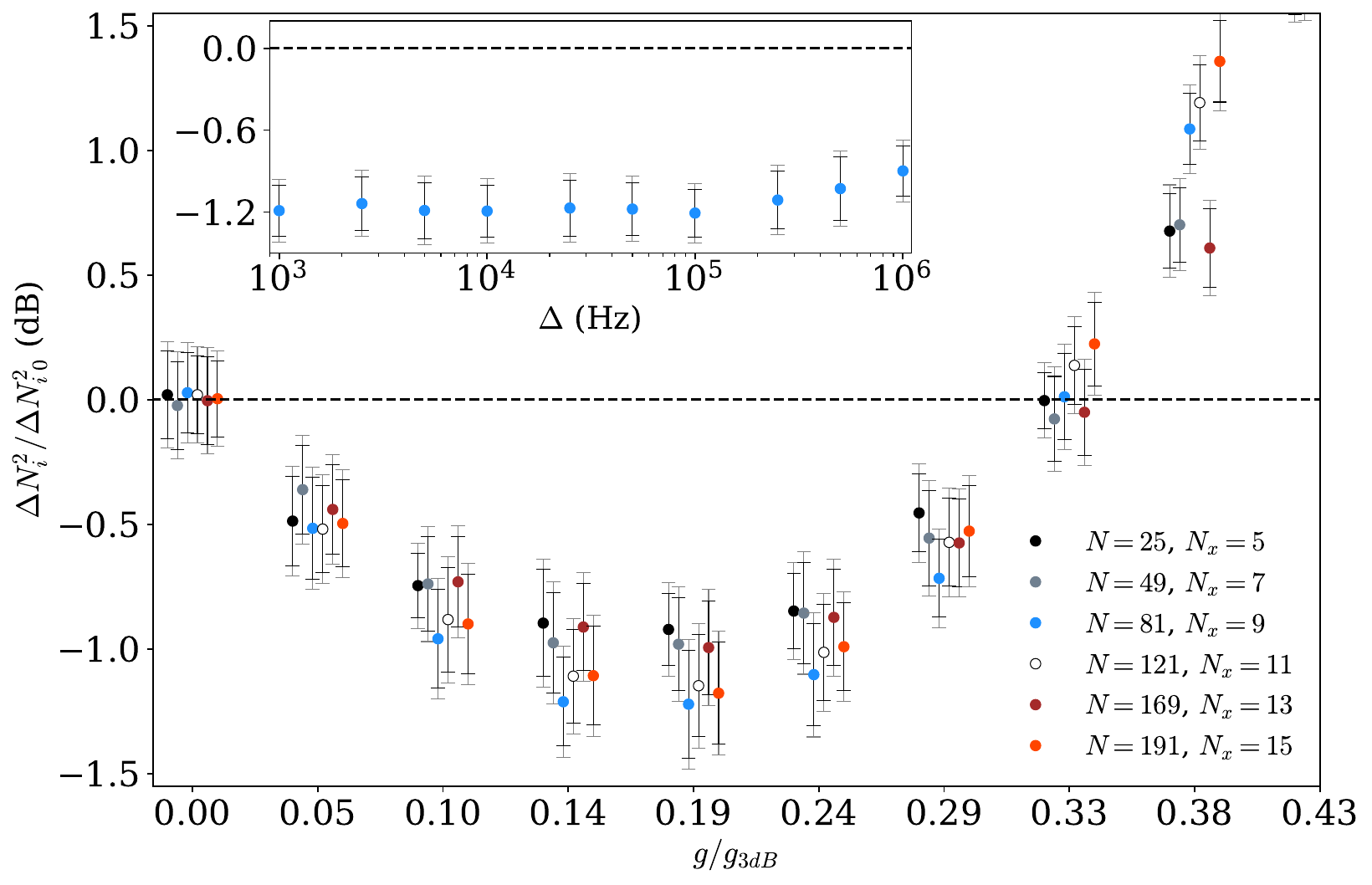}
    \caption{
    Minimum normalized nullifier variance $\Delta N_i^2 / \Delta N_{i,0}^2$ as a function of the pump strength $g/g_{3\mathrm{dB}}$ for increasing lattice sizes $(N, N_x)$. Error bars denote experimental uncertainty; the dashed line marks the vacuum level. 
    Inset: nullifier variance versus frequency spacing $\Delta$ at the optimal pump power.
    }

    \label{fig:Null_N}
\end{figure}

Figure~\ref{fig:Null_SqHC}b) shows the nullifier variance as a function of the pump power $g$, normalized to the value $g_{3\mathrm{dB}}$ at which we obtain 3~dB of gain when the JPA is operated as a non-degenerate parametric amplifier, with a single pump tone.  
We see a minimum in the nullifier squeezing for $g/g_{3\text{dB}}=0.19$.
For higher pump powers the squeezing progressively decreases. 
This behavior is consistent with the presence of losses in the parametric oscillator. 
We performed several experiments with different pumping schemes, altering the cluster state width $N_x$ and the number of measured modes $N$. 
The results were consistent for all $N$ and $N_x$ tested, with the same level of nullifier squeezing and the same behavior of the nullifier as a function of pump power, as shown for the square lattice case in Fig.~\ref{fig:Null_N}.
We also studied the dependence of the nullifier on measurement bandwidth, or spacing between tones in the comb $\Delta=\tfrac{2 \pi}{T}$. 
We performed a series experiments at optimal squeezing power $g/g_{3\text{dB}}=0.19$ for the case of the square lattice with $N=81$ and $N_x=9$ and the measurement bandwidth in the range $\Delta\in[1~\text{kHz},1~\text{MHz}]$. 
We see in the inset to Fig.~\ref{fig:Null_N} that the nullifier squeezing remains constant as a function of $\Delta$, with a slight degradation at large delta.
We attribute this degradation to weaker entanglement between modes that are further from $\omega_0$.

\section{Hidden Entanglement Analysis}

The presence of off-diagonal elements in $\matr{U}$ provides a direct diagnostic of hidden entanglement~\cite{menicucci_graphical_2011}. In our system the hidden entanglement arises from higher order mixing processes and takes the form of pairs of off-diagonals.
At sufficient pumping amplitude, these mixing processes appear above the noise floor of the measured matrix $\matr{U}$.   
To quantify the diagonality of $\matr{U}$, we define the \emph{Hidden Entanglement Ratio} (HER) 
\begin{equation}
\mathrm{HER} = \sum_{i,k\in K}\frac{N}{N_{k}}\frac{| U_{i,i+k
}|}{\text{Tr}[\matr{U}]}.
\label{eq:HER}
\end{equation}
which quantifies the ratio between the average values of the sum of off-diagonals and the main diagonal of $\matr{U}$.
Here $N_{k}$ is the number of non-zero elements in the $k$-th off-diagonal, and $K$ is the set containing the index of non-zero off-diagonals: $K = \{\pm2, \pm2N_x\}$ for square lattice and $K =
\{\pm2,\pm(N_x-2),\pm N_x\}$ for honeycomb. 
We restrict the sum on the numerator to the expected non-zero off-diagonal elements to limit fluctuation of HER at low $g$ due to the noise floor (i.e. signal and noise scaling linearly with $N$). 
We see that this definition of HER well describes the diagonality of $\matr{U}$.

\begin{figure*}
    \centering
    \includegraphics[width=\textwidth]{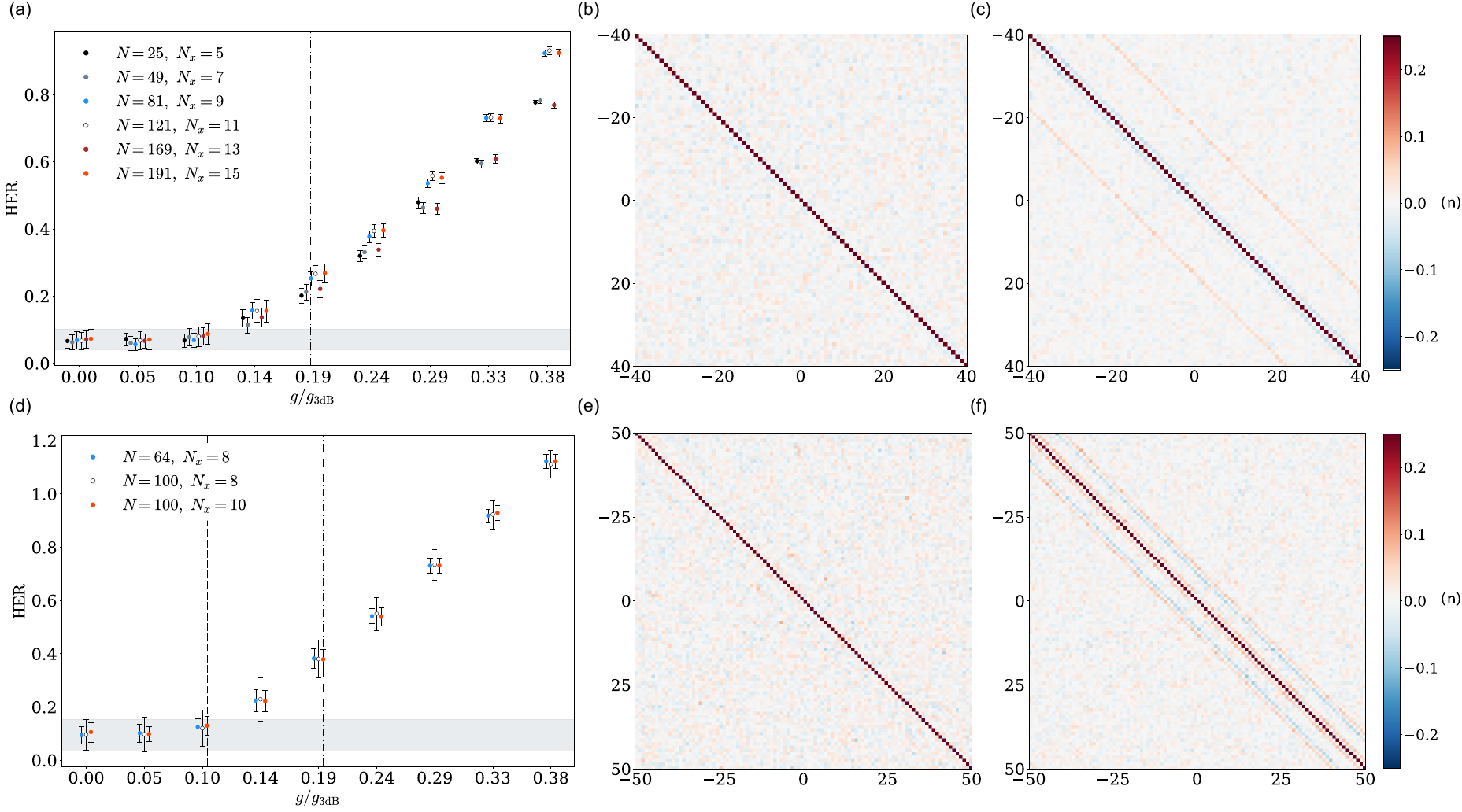}
    \caption{
    Hidden entanglement analysis.
    (a) Hidden entanglement ratio (HER) versus normalized pump amplitude $g/g_{3\mathrm{dB}}$ for the square lattice; the shaded area indicates the low-HER regime.
    (b),(c) Representative $\matr{U}$ matrices at low pump power and at optimal squeezing, respectively.
    (d) HER versus $g/g_{3\mathrm{dB}}$ for the honeycomb lattice.
    (e),(f) Corresponding $\matr{U}$ matrices at low pump power and at optimal squeezing.
    }
    \label{fig:hidden_entanglement}
       
\end{figure*}

Figure~\ref{fig:hidden_entanglement} shows the measured HER with two samples of the matrices $\matr{U}$ at different pump amplitudes, for both square lattice (upper panels) and honeycomb lattices (lower panels). 
Panels a) and d) show the HER as a function of normalized pump amplitudes $g/g_{3\mathrm{dB}}$ and different lattice sizes.
At low pump amplitude HER remains below the noise floor. For higher values of $g/g_{3\mathrm{dB}}$ some hidden correlations start to arise above the noise floor.  
In Fig.~\ref{fig:hidden_entanglement}b) and c) we plot matrix $\matr{U}$ for the square lattice at $g/g_{3\mathrm{dB}}$ corresponding to a nullifier squeezing of $-1$~dB (dashed line in panel a), and  maximum squeezing (dot-dashed line in panel a). Fig.~\ref{fig:hidden_entanglement}e) and f) show the same for the honeycomb lattice.

Overall we see a signature of hidden entanglement at optimal squeezing power and no apparent hidden entanglement up to $\sim -1$~dB of squeezing. 
These correlations are the result of $x$-quadrature correlations between modes outside the target graph, corresponding to a weak connection to second nearest neighbors in the graph. They appear as a weak signal in the covariance matrix and they are amplified in the $\matr{U}$. 
At optimal squeezing they are a factor of $5$ weaker than the dominant canonical correlations.
We estimate the ratio between the desired canonical correlations $c_{\text{can}}$ and hidden correlations $c_{\text{hid}}$ to be $\sim 0.208 \pm 0.052$.
Here, $c_{\text{can}}$ and $c_{\text{hid}}$ are obtained by averaging the absolute values of the covariance matrix elements corresponding to desired and hidden correlations respectively. 
We observe a higher total hidden entanglement for the honeycomb, due to the fact that the asymmetric pumping scheme shown in Fig.~\ref{fig:CVCS_HC}a) is not as effective at canceling the second-order idlers in comparison with the square lattice case. 

\section{Conclusions}

In this work we have demonstrated the first experimental realization of two-dimensional continuous-variable cluster states in the microwave domain, achieving 2D connectivity between 191 frequency modes traveling into a transmission line. 
By carefully engineering the frequencies, amplitudes, and phases of multiple parametric pump tones applied to a Josephson Parametric Amplifier, we generated both square and honeycomb lattice topologies with a single experimental platform. 

The cluster states were verified through a nullifier test, reaching up to -1.2~dB of squeezing below the vacuum level, consistent across all lattice sizes and frequency spacing tested. 
This result extends the frontier of microwave CV cluster state generation beyond the one-dimensional square-ladder topology of our previous work~\cite{lingua_continuous-variable_2025}, and represents an important step toward scalable continuous-variable measurement-based quantum computing in superconducting systems.

A central figure of merit beyond nullifier squeezing is the degree of hidden entanglement, quantified through the Hidden Entanglement Ratio. 
We find that hidden entanglement remains below the noise floor up to nullifier squeezing of approximately -1~dB, and is present but well-controlled at maximum squeezing -1.2~dB, where hidden correlations are a factor of five weaker than the dominant canonical correlations. 
This demonstrates that the engineered pump interference scheme is effective at suppressing unwanted correlations.

An open question regards the extent to which interference induced by additional pumps can suppress hidden entanglement.  
We applied the method developed in~\cite{cortinovis_inverse_2025} to find a multifrequency pump waveform which achieves the target covariance matrix of the square and honeycomb lattices, without hidden entanglement. 
While this method did accurately recover the components of the pump waveform which we presented above, we found that the additional pump frequency components found by the method, did not suppress hidden entanglement as measured by HER.  
It remains to be seen if hidden entanglement can be suppressed through more elaborate pump engineering.

\section{Data availability}
Data available from the corresponding author upon reasonable request.

\section{Acknowledgments}
We acknowledge Joe Aumentado at the National Institute of Standards and Technology (NIST) for helpful discussions and for providing the JPA used in this experiment. 
This work was partially supported by the Knut and Alice Wallenberg Foundation through the Wallenberg Center for Quantum Technology (WACQT).

\section{Author contributions}
F. L. and M. C. performed both the experiments and theoretical analysis. F. L. derived the pumping schemes for square and honeycomb lattices. M. C. performed the analysis of the hidden entanglement. J. C. R. H. performed the calibration of the experimental setup. F. L. and M. C. prepared the first draft of the manuscript. D. B. H. supervised the work. All the authors discussed, reviewed and edited the manuscript.

\section{Competing interests}
D. B. H. is part owner of the company Intermodulation Products AB, which produces the digital microwave platform used in this experiment.

\appendix
\section{Experimental Setup}\label{ExpSetup}

\begin{figure}
    \centering
    \includegraphics[width=\columnwidth]{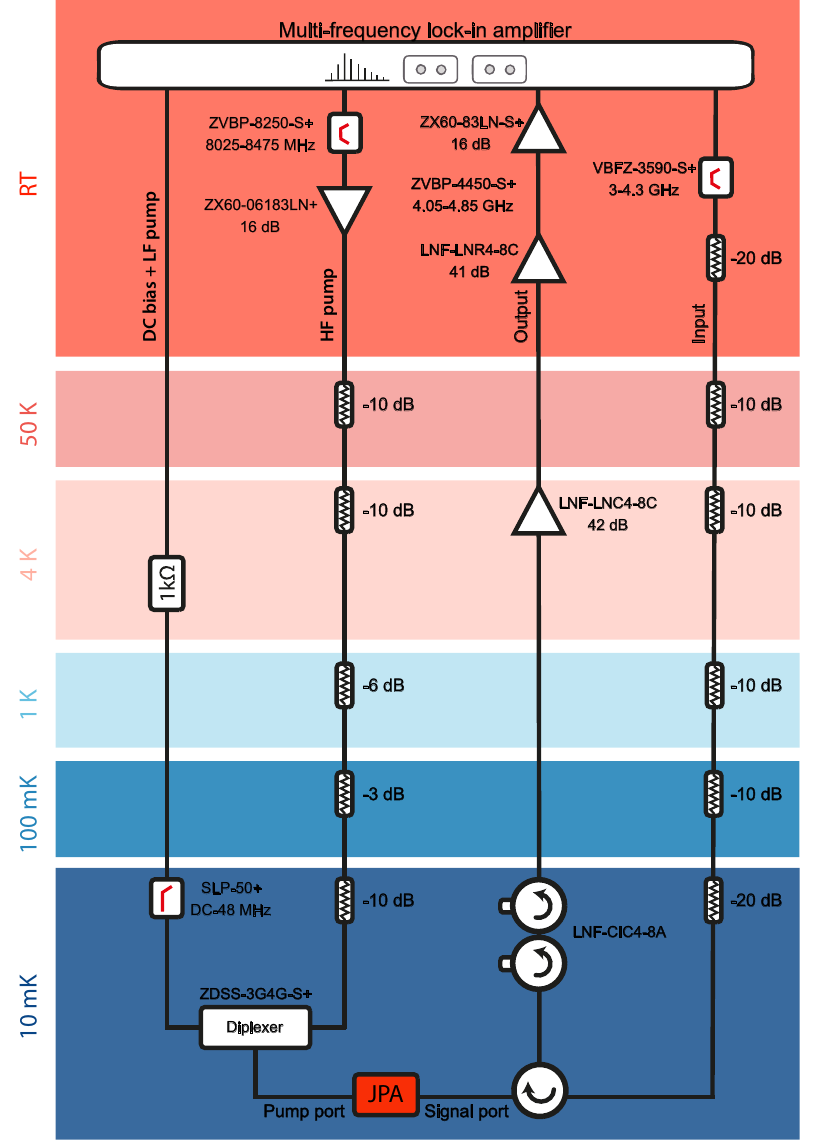}
    \caption{ Schematic of the experimental setup, including room temperature and cryogenic electronics.}
    \label{fig:setup}
\end{figure}

A schematic of the measurement setup is shown in Fig.~\ref{fig:setup}.  A Josephson Parametric Amplifier (JPA) placed at the bottom of a dilution refrigerator (DR) and thermalized at $10$~mK. The resonance frequency of the JPA is dc-flux biased to around $\omega_0\simeq4.2$~GHz. The multifrequency pump signal is combined with this DC flux bias using a diplexer, and applied to the pump port of the JPA.  

Cold attenuators and various temperature stages on the input line absorb thermal blackbody photons from higher temperatures, ensuring that the JPA mode sees the vacuum fluctuations of the 50$\Omega$ input line.  The circulator directs the vacuum fluctuations to the signal port of JPA.  The multifrequency pumped JPA correlates these vacuum fluctuations upon reflection, giving it the desired covariance between frequency modes.  This entangled radiation is directed through the circulator to the output line, where two isolators at $10$~mK protect the JPA from the noise added by the following amplifiers.  Superconducting coax is used to between $10$~mK and $4$~K on the output line to eliminate signal loss.  

The noise floor in the measurement is set by the first cryogenic low-noise amplifier at  $4$~K, which typically adds between 12 and 16 photons to each frequency mode, far above the vacuum level. This added noise is however completely uncorrelated between frequency modes, and therefore does not effect the off-diagonal elements of the covariance matrix.  The added noise must be subtracted from the diagonal elements, requiring careful calibration of both the added noise level, and gain in the amplification chain. See supplemental material~\cite{lingua_supplemental_2026} for a detailed discussion of the calibration procedure.

The amplified signal is demodulated by a multifrequency lockin amplifier, which also supplies the multi-frequency pump waveform. 
The multi-frequency lock-in amplifier is actually firmware running on a microwave digital platform called Presto created from a radio-frequency system-on-a-chip~\cite{tholen_measurement_2022}.
This instrument also supplies the DC bias for the JPA. 
All signals across all output ports and all input ports are synchronized through one master clock, and all frequencies in the pump waveform, and of the frequency modes, are tuned to be integer multiples of the measurement bandwidth, $\Delta$.  
This condition ensures one common phase references across all frequencies in the measurement.  An additional phase correction is required to account for propagation delay and latency in the high-speed data converters. We measure this correction to be $1.89$~rad/MHz.

\section{Covariance Matrix Measurement}\label{app_CovMeas}

We measure the output voltage noise quadratures $x_i, p_i$ (in Volts) at each frequency within our mode basis. We build the experimental covariance matrix $\matr{V}^\text{M}$ in units of photon number:
\begin{equation}
    V^\text{M}_{ij} =  \dfrac{ \expval{\text{cov}[\vec{q}-\expval{\vec{q}\,} ]_{ij} }_M}{Z_c \hbar \Delta \sqrt{\omega_i \omega_j}}.
    \label{eq:cov-mat}
\end{equation}
Where $\langle\cdot,\cdot\rangle_M$ denotes the statistical average between $10^6$ measurements (i.e. time windows), and $Z_c=\SI{50}{\ohm}$ is the impedance seen at the signal port. Based on our indexing convention, the central qumode $i=0$ is positioned at the frequency $\omega_0$.

To isolate the purely quantum covariance matrix $\matr{V}^\text{quant}$ of the output qumodes, it is necessary to subtract the uncorrelated classical noise introduced by the HEMT at 4K from the diagonal of $\matr{V}^\text{M}$ ~\cite{jolin_multipartite_2023,lingua_continuous-variable_2025}. 
This subtraction process relies on a precise calibration of the amplification chain~\cite{mariantoni_planck_2010}. Once the classical noise is removed, we must ensure that the covariance matrix represents a physically valid state that obeys the Heisenberg uncertainty principle. We achieve this by applying a constrained minimization technique~\cite{shchukin_recovering_2016}, which ultimately yields the legitimate physical covariance matrix $\matr{V}$. 
Interestingly, we noticed no measurable difference in the computed nullifiers with or without the application of the constrained minimization. 
Additional details regarding both the calibration and the reconstruction procedure can be found in the Supplemental Material~\cite{lingua_supplemental_2026}.

The rotation of all the quadratures by the same angle $\theta$ shown in the nullifier results corresponds to the rotation of the covariance matrix through the unitary transformation:
\begin{equation}
    \matr{V}^\prime = \matr{U}_\theta \cdot \matr{V}\cdot \matr{U}_\theta^{T}
    \label{eq:Vrot}
\end{equation}
where $\matr{U}_\theta=\mathbb{I}_N\otimes \matr{R}_{\theta}$, $N$ is the number of modes and 
\begin{equation}
    \matr{R}_\theta=\begin{pmatrix}
    \cos\theta & -\sin\theta\\
    \sin\theta & \cos\theta
    \end{pmatrix},
\end{equation}
Note that rotating the covariance matrix is equivalent to rotating the quadrature vector $\vec{q}\,^\prime=\matr{R}_\theta\cdot\vec{q}$.

\bibliographystyle{sn-aps}   
\bibliography{Refs}             
\nocite{pfister_continuous-variable_2020, malnou_three-wave_2021, holevo_evaluating_2001, weedbrook_gaussian_2012, petrovnin_generation_2023, mariantoni_planck_2010, clerk_introduction_2010, tholen_measurement_2022, simon_quantum-noise_1994, weedbrook_gaussian_2012, shchukin_recovering_2016, yamamoto_principles_2016}

\end{document}


\preprint{APS/123-QED}

\title{Supplemental material: Continuous-variable two-dimensional cluster states in the microwave domain} 

\author{Fabio Lingua}
 \email{lingua@kth.se}
 \affiliation{Department of Applied Physics, KTH Royal Institute of Technology, SE-10691 Stockholm, Sweden}

\author{Michele Cortinovis}
 \affiliation{Department of Applied Physics, KTH Royal Institute of Technology, SE-10691 Stockholm, Sweden}
 \affiliation{Dipartimento di Fisica, Politecnico di Milano, I-20133 Milano, Italy}
 
\author{J. C. Rivera Hernández}
 \affiliation{Department of Applied Physics, KTH Royal Institute of Technology, SE-10691 Stockholm, Sweden}

\author{David B. Haviland}
 \affiliation{Department of Applied Physics, KTH Royal Institute of Technology, SE-10691 Stockholm, Sweden}

\date{\today}

\maketitle

\section{Cluster States and Nullifiers}

Continuous-variable (CV) cluster states are multi-partite entangled states defined by a canonical graph between $n$ \emph{qumodes}.
These states are continuous superpositions of $n$-mode position states with a CV ${C_{z, ij}=e^{iA_{ij} \hat{x}_i \hat{x}_j}}$ gate applied between mode pairs corresponding to the edges of the canonical graph:
\begin{multline}
    \ket{\Psi}=C_\text{z}\ket{0}_{p_1}\ket{0}_{p_2}\dots\ket{0}_{p_n}=\\
    =\frac{1}{2\pi}\iint d\vec{x} \prod_{i,j\in G} e^{iA_{ij} x_i x_j} \ket{x_1}\ket{x_2}\dots\ket{x_n},
    \label{CS_def}
\end{multline}
where $\ket{0}_{p_1}\ket{0}_{p_2}\dots\ket{0}_{p_n}$ is the vacuum in the momentum basis, $d\vec{x} \equiv dx_1 dx_2\dots dx_n$, and
$C_\text{z}=\prod_{i,j\in G} C_{z,\, ij}$ contains the sequence of pairwise $C_z$ gate correlations describing the canonical graph $G$.

\begin{figure}
    \includegraphics[width=\columnwidth]{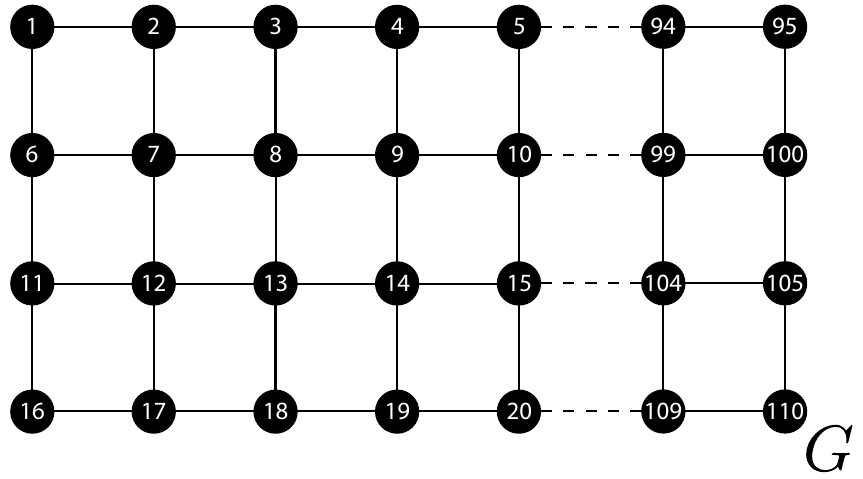}
    \caption{\label{fig:CState} Example of a square-ladder cluster state. Links between nodes correspond to CV $C_z$ gate operations. 
    }
\end{figure}

To realize quantum computation the canonical graph should not be fully connected. 
A graph for measurement-based quantum computation (MBQC) typically features nearest-neighbor connectivity and a lattice topology (see Fig.~\ref{fig:CState} for an example).
Cluster states are defined by their symmetry properties, described by the multiplicative stabilizer group $\mathcal{S}$.
For a stabilizer operator $S\in\mathcal{S}$, the cluster state $\ket{\Psi}$ is invariant under transformation by $S$: 
\begin{equation}
    S\ket{\Psi}=\ket{\Psi}, 
    \label{Stab_eq}
\end{equation}
i.e. $\ket{\Psi}$ is an eigenstate of $S$ with eigenvalue $1$. $S$ is in general defined in terms of local stabilizers $S_i$
\begin{equation}
S= \bigoplus_{i=1}^n S_i,
\end{equation}
where the direct sum on $i$ runs over all the number of modes.
Local stabilizers $S_i$ obey~\eqref{Stab_eq} and take the form
\begin{equation}
    S_i=X_i\bigotimes_{j\in G_i} Z_j, \label{Sdef}
\end{equation}
where $X_i=e^{-i\xi_i \hat{p}_i}$, $Z_j=e^{i\varpi_j \hat{x}_j}$ are the CV analogous of Pauli gates~\cite{pfister_continuous-variable_2020}, $\xi_i,\varpi_j\in\mathbb{R}$ and $G_i$ the graph of modes connected to $i$. 
Note that in Eq.~\eqref{Sdef} the quadrature operators $\hat{p}_i$, $\hat{x}_j$ commute for $i\neq j$. 
One can therefore treat the product of exponential operators as the exponential of the sum.
It follows for Eq.~\eqref{Stab_eq} to be satisfied,
\begin{equation}
    -i\xi_i\left[\hat{p}_i - \sum_{j\in G_i}\frac{\varpi_j}{\xi_i} \hat{x}_j\right]\ket{\Psi}=0\,. \label{Stab_Null}
\end{equation}
From~\eqref{Stab_Null} we define the \emph{nullifier} for mode $i$ as
\begin{equation}
    N_i=\hat{p}_i - \sum_{j\in G_i}h_{ij} \hat{x}_j\,,
\end{equation}
where $h_{ij}=\frac{\varpi_j}{\xi_i}\in\mathbb{R}$.
A suitable choice of $h_{ij}$ fixes the form of the CV quantum gates and yields the proper stabilization of $\ket{\Psi}$ (see the section below for details).
In the limit of infinite squeezing, the expectation values and variance of the nullifier operators satisfy the relations
\begin{eqnarray}
    &\expval{N_i}&=\bra{\Psi}N_i\ket{\Psi} = 0\,,\\
    &\Delta N_i^2 &= \bra{\Psi}(N_i - \expval{N_i})^2\ket{\Psi} = 0\,.
    \label{DNi2}
\end{eqnarray}
For finite squeezing it is sufficient to show that $\expval{N_i}\simeq 0$, and $\Delta N_i^2<\Delta N_{i\;0}^2 $, where $\Delta N_{i\;0}^2=\bra{0}(N_i - \expval{N_i})^2\ket{0}$ is the vacuum fluctuations of the nullifier, i.e. the variance of $N_i$ computed for the vacuum, or in the absence of the entangling pumps.    
In our experiment we compute the squeezing of the nullifiers in dB following the standard definition $10\log_{10}( \Delta N_i^2/\Delta N_{i\;0}^2) $.
The nullifier test is considered passed if the ratio $\Delta N_i^2/\Delta N_{i\;0}^2$ is below \SI{0}{\dB} for all modes $i\in G$.

Explicitly, Eq.~\eqref{DNi2} reads
\begin{multline}
    \Delta N_i^2 = \Delta \hat{p}_i^2 + \sum_{jk}\Delta \hat{x}_j\Delta\hat{x}_k +\\ 
    -\sum_{j}\left(\Delta\hat{p}_i\Delta \hat{x}_j + \Delta \hat{x}_j\Delta\hat{p}_i\right) .
    \label{eq:DNi2_xplct}
\end{multline}
Equation~\eqref{eq:DNi2_xplct} is readily evaluated from the elements of the covariance matrix $V$.

\subsection{Cluster State Stabilization Condition}
To prove the stabilization condition of $\ket{\Psi}$, one has to prove that there exists a stabilizer $S_i$ $\forall i\in G$.
Explicitly writing $S_i$ acting on the definition of the cluster state~\eqref{CS_def} and using~\eqref{Sdef} one obtains 

\begin{widetext}
\begin{multline}
    S_i\ket{\Psi}=S_iC_\text{z}\ket{0}_{p_1}\dots\ket{0}_{p_i}\dots\ket{0}_{p_n} = \frac{1}{2\pi}\iint d\vec{x}\; e^{-i\xi_i\left[\hat{p}_i - \sum\limits_{j\in G_i}h_{ij} \hat{x}_j\right]}\prod_{k,j\in G} e^{iA_{kj} x_k x_j} \ket{x_1}\dots\ket{x_i}\dots\ket{x_n}=\\
    =\frac{1}{2\pi}\iint d\vec{x}\; e^{-i\xi_i\left[\hat{p}_i - \sum\limits_{j\in G_i}h_{ij} \hat{x}_j\right]}\prod_{k=i,j\in G_i}e^{iA_{ij} x_i x_j}\prod_{k,j\in G-G_i} e^{iA_{kj} x_k x_j} \ket{x_1}\dots\ket{x_i}\dots\ket{x_n}=\\
    =\frac{1}{2\pi}\iint d\vec{x}\; e^{-i\xi_i\hat{p}_i} e^{i\left[\xi_i\sum\limits_{j\in G_i}h_{ij} x_j + \sum\limits_{j\in G_i}A_{ij} x_i x_j\right]}\prod_{k,j\in G-G_i} e^{iA_{kj} x_k x_j} \ket{x_1}\dots\ket{x_i}\dots\ket{x_n},
    \label{Stab_Psi}    
\end{multline}
\end{widetext}
where we exploited the fact that $\ket{x_j}$ is eigenstate of $\hat{x}_j$ with eigenvalue $x_j$, and used the properties $e^{i\beta\hat{x}_j}\ket{x_j}=e^{i\beta x_j}\ket{x_j}$ and $[x_k, x_j]=0$ $\forall j,k$.
The momentum operator in the exponent is a generator of translation $e^{-i\xi_i\hat{p}_i}\ket{x_i}=\ket{x_i + \xi_i}$ from which it follows
\begin{widetext}
\begin{equation}
    \frac{1}{2\pi}\iint d\vec{x}\;  e^{i\left[\xi_i\sum\limits_{j\in G_i}h_{ij} x_j + \sum\limits_{j\in G_i}A_{ij} x_i x_j\right]}\prod_{k,j\in G-G_i} e^{iA_{kj} x_k x_j} \ket{x_1}\dots\ket{x_i + \xi_i}\dots\ket{x_n}.
    \label{Stab_Psi1}    
\end{equation}
\end{widetext}
Changing the variable of integration $x_i\longrightarrow x_i - \xi_i$ results in
\begin{widetext}
\begin{multline}
    \frac{1}{2\pi}\iint d\vec{x}\;  e^{i\sum\limits_{j\in G_i}\left[\xi_i h_{ij} x_j + A_{ij}(x_i - \xi_i) x_j\right]}\prod_{k,j\in G-G_i} e^{iA_{kj} x_k x_j} \ket{x_1}\dots\ket{x_i }\dots\ket{x_n}=\\
    =\frac{1}{2\pi}\iint d\vec{x}\;  e^{i\sum\limits_{j\in G_i}\xi_i\left( h_{ij} - A_{ij}\right)x_j}\cdot e^{i\sum\limits_{j\in G_i} A_{ij}x_i x_j}\prod_{k,j\in G-G_i} e^{iA_{kj} x_k x_j} \ket{x_1}\dots\ket{x_i }\dots\ket{x_n}=\\
    =\frac{1}{2\pi}\iint d\vec{x}\;  e^{i\sum\limits_{j\in G_i}\xi_i\left( h_{ij} - A_{ij}\right)x_j}\cdot \prod_{k,j\in G} e^{iA_{kj} x_k x_j} \ket{x_1}\dots\ket{x_i }\dots\ket{x_n}.
    \label{Stab_Psi2}    
\end{multline}
\end{widetext}
For Eq.~\eqref{Stab_Psi2} to be equal to Eq.~\eqref{CS_def} one requires
\begin{equation}
     e^{i\sum\limits_{j\in G_i}\xi_i\left( h_{ij} - A_{ij}\right)x_j}=1,
     \label{cond_large}
\end{equation}
which is certainly satisfied for
\begin{equation}
    h_{ij} = A_{ij}.
    \label{cond_stringent}
\end{equation}
Conditions~\eqref{cond_large}~and~\eqref{cond_stringent} essentially prove the symmetry transformation $S_i\ket{\Psi}=\ket{\Psi}$.
The particular choice of $h_{ij}$  satisfying~\eqref{cond_stringent} fixes the definition of gates $X$, $Z$ in the MBQC toolbox. 

\begin{figure*}[t]
    \centering
    \includegraphics[width=\linewidth]{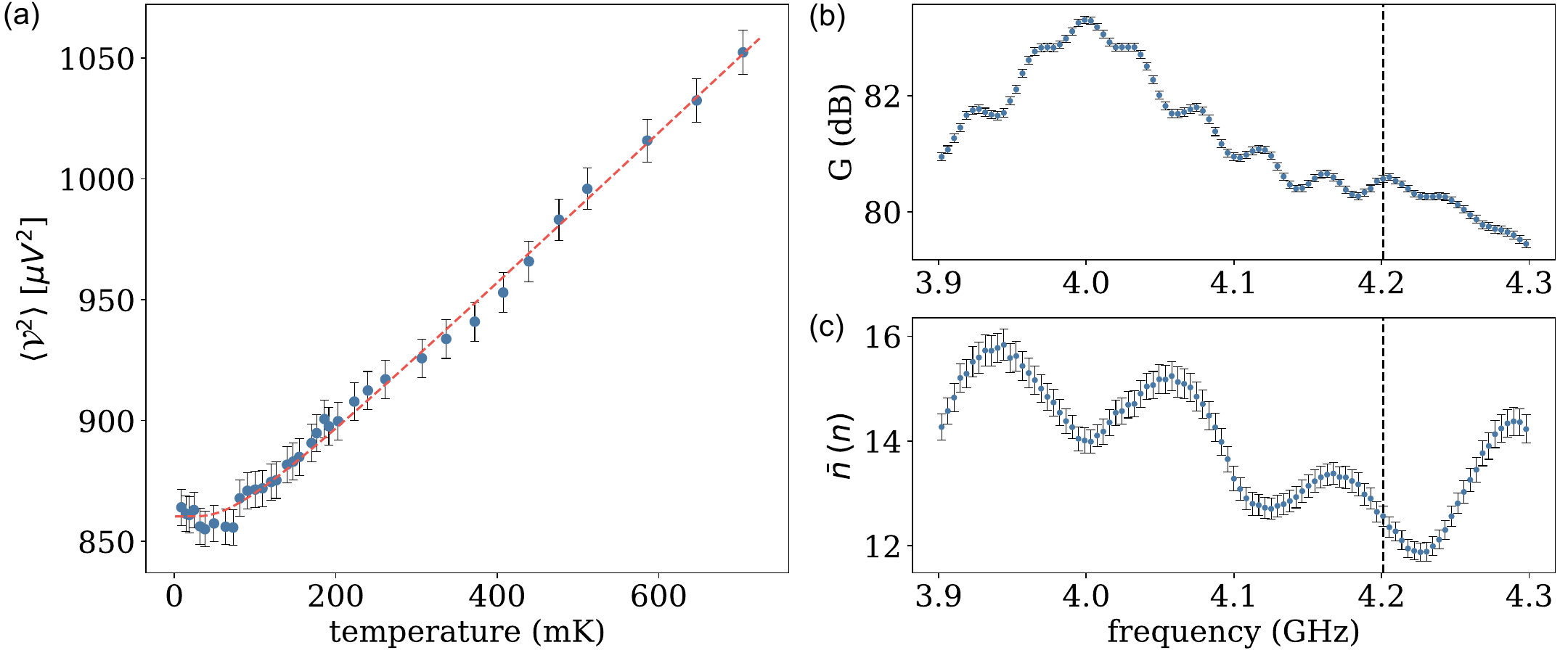}
    \caption{(a) Variance of the voltage noise at \SI{4.2}{\giga\hertz} measured as a function of temperature. The red dashed line corresponds to the fit to Eq.~\eqref{eq:planck-v}. Such a fit at each frequency provides the gain $G$ and the added number of noise photons $\bar{n}$, plotted as a function of frequency in (b) and (c) respectively. The black error bars in all figures refer to the measurement uncertainty.}
    \label{fig:planck}
\end{figure*}

\section{Covariance Matrix Reconstruction}
To analyze the quantum correlations induced by the JPA we must remove the uncorrelated classical noise added by our amplification chain.
This chain acts as a noisy bosonic Gaussian channel~\cite{malnou_three-wave_2021}, where each mode $i$ experiences a frequency-dependent gain $G(\omega_i)$ and an average number of added noise photons $\bar{n}_i$.
The effect of a bosonic Gaussian channel on an arbitrary Gaussian state transforms the mean and covariance matrix as follows~\cite{holevo_evaluating_2001, weedbrook_gaussian_2012}
\begin{equation}
    \bar{x} \rightarrow T\bar{x} + d, \;\;\;\;\; V \rightarrow TVT^T + N,
    \label{eq:Gaussian-channel}
\end{equation}
where $T=\bigoplus_i \sqrt{G(\omega_i)}I_i$ and $N = \bigoplus_i \left(G(\omega_i) - 1\right)(\bar{n}_i + 1/2)I_i$.
Since $T$ and $N$ are diagonal matrices, the resulting measured covariance matrix $V^\text{meas}$ is
\begin{equation}
    V^\text{meas} = T^2 V^\text{quant} + N,
    \label{eq:V-trans}
\end{equation}
where $V^\text{quant}$ represents the covariance matrix of the quantum state scattered by the JPA before amplification.

Unavoidable uncertainties in estimating the added noise $N$ result in unphysical covariance matrices due to excessive noise subtraction.
To circumvent this problem we exploit the fact (verified by experiment) that the amplification chain's added classical noise at different frequencies is completely uncorrelated.
This added noise therefore only contributes to the diagonal elements of $V^\text{meas}$.
In the absence of pumps, the ratio between the quantum noise and the added classical noise on the diagonal is
\begin{equation}
    \dfrac{V^\text{quant}_{0\,ii}}{V^\text{classic}_{0\,ii}} = \dfrac{1}{2\bar{n}_i + 1}.
    \label{eq:ratio-Vqc-off}
\end{equation}
The measured covariance matrix with the pump off $V^\text{meas}_0$ after amplification is given by
\begin{equation}
    V^\text{meas}_{0\,ii} = G(\omega_i) \left(V^\text{quant}_{0\,ii} + V^\text{classic}_{0\,ii}\right).
    \label{eq:Vm0-diag}
\end{equation}
Using Eqs.~\eqref{eq:ratio-Vqc-off}~and~\eqref{eq:Vm0-diag}, and assuming that the classical uncorrelated noise remains the same for the pump-on case, we recover the quantum covariance matrix as~\cite{petrovnin_generation_2023}: 
\begin{equation}
    V^\text{quant} = T^{-2} \left( V^\text{meas} - V^\text{meas}_0 \right) + V^\text{quant}_0, 
    \label{eq:Vq-rec}
\end{equation}
where $V^\text{quant}_0$ can be calculated from $V^\text{meas}_0$ as
\begin{align}
    V^\text{quant}_{0\,ij} = 
    \begin{cases} 
        \dfrac{1}{2\bar{n}_i+2} \; G(\omega_i)^{-1} \; V^\text{meas}_{0\,ij}, \;\; \text{for} \;\; i=j \\ \\
        G(\omega_i)^{-1} \; V^\text{meas}_{0\,ij}, \;\; \text{for} \;\; i \neq j
    \end{cases}
    \label{eq:Vqoff}
\end{align}

We estimate $G(\omega_i)$ and $\bar{n}_i$ by measuring the Johnson-Nyquist noise spectral density emitted by a \SI{50}{\ohm} source (the isolator) as a function of temperature $T_\mathrm{mxc}$.
This method, known as Planck spectroscopy~\cite{mariantoni_planck_2010}, involves slowly heating the mixing chamber to vary the isolator's temperature and allowing \SI{2}{\hour} for thermalization before measuring.
The amplified power spectral density of the noise is given by~\cite{clerk_introduction_2010}
\begin{equation}
    P = \dfrac{1}{2} G(\omega_i) hf_i \left[ \coth{\left( \dfrac{hf_i}{2k_B T_\mathrm{mxc}} \right)} + 2\bar{n}_i + 1 \right].
    \label{eq:planck}
\end{equation}
In terms of the voltage variance over a measurement bandwidth $\Delta$, we have
\begin{equation}
    \langle \mathcal{V}^2 \rangle = 4 \Delta P \times \SI{50}{\ohm}.
    \label{eq:planck-v}
\end{equation}
Figure~\ref{fig:planck} shows the fit of Eq.~\eqref{eq:planck-v} to the measured voltage variance. 
We measure at 192 frequencies simultaneously using the multifrequency lock-in~\cite{tholen_measurement_2022}. 
Performing this fit at each frequency, we extract the gain $G(\omega_i)$ and added noise $\bar{n}_i$ as a function of frequency, as shown in Fig.~\ref{fig:planck}b and Fig.~\ref{fig:planck}c, respectively.

Inherent difficulties associated with extracting the quantum noise from the added noise, as well as uncertainties in our calibration, lead to a reconstructed $V^\text{quant}$ which is nonphysical, in the sense that it violates the Heisenberg uncertainty principle. 
The uncertainty principle for a covariance matrix expressed in photon number ($\Delta x \Delta p \geq 1/2$ or $\hbar=1$) is stated as~\cite{simon_quantum-noise_1994, weedbrook_gaussian_2012}
\begin{equation}
    V^\text{quant} \pm \dfrac{i}{2} \Omega \geq 0, \;\;\;\; \text{with} \;\; \Omega = \bigoplus_i \begin{pmatrix}
                    0 & 1\\
                    -1 & 0
                \end{pmatrix}.
    \label{eq:physicality}
\end{equation}

To compute the best-approximated physical covariance matrix $V$, we perform a constrained minimization of an objective function~\cite{shchukin_recovering_2016}
\begin{equation}
    \min_{V} \left( \max_{ij}\frac{ \left|V^\text{quant}_{ij} - V_{ij} \right|}{\sigma^{V^\text{quant}}_{ij}} \right),
    \label{eq:shchukin}
\end{equation}
where $\sigma^{V^\text{quant}}_{ij}$ represents the experimental error associated with the $ij$-th element of $V^\text{quant}$.

\section{Error Analysis}
To compute the best approximated physical covariance matrix and estimate the uncertainty in the nullifiers, we need the experimental errors in the quantum covariance matrix $\sigma^{V^\text{quant}}_{ij}$.
We estimate these errors by combining the uncertainties in the calibration and measurement through error propagation of Eq.~\eqref{eq:Vq-rec}.
The uncertainty of each element of $V^\text{quant}$ is given by:
\begin{widetext}
\begin{multline}
    \sigma^{V^\text{quant}}_{ij} = \sqrt{\left( \dfrac{\partial V^\text{quant}_{ij}}{\partial G(\omega_i)} \right)^2 \sigma_{G(\omega_i)}^2 + \left( \dfrac{\partial V^\text{quant}_{ij}}{\partial V^\text{meas}_{ij}} \right)^2 \sigma_{V^\text{meas}_{ij}}^2 + \left( \dfrac{\partial V^\text{quant}_{ij}}{\partial V^\text{meas}_{0\,ij}} \right)^2 \sigma_{V^\text{meas}_{0\,ij}}^2 + \left( \dfrac{\partial V^\text{quant}_{ij}}{\partial V^\text{quant}_{0\,ij}} \right)^2 \sigma_{V^\text{quant}_{0\,ij}}^2} = \\
    = \sqrt{\left| \dfrac{(V^\text{meas}_{ij} - V^\text{meas}_{0\,ij})}{G(\omega_i)^{2}} \right|^2 \sigma_{G(\omega_i)}^2 + G(\omega_i)^{-2} \left( \sigma_{V^\text{meas}_{ij}}^2 + \sigma_{V^\text{meas}_{0\,ij}}^2 \right) + \sigma_{V^\text{quant}_{0\,ij}}^2},
    \label{eq:error-Vq}
\end{multline}
\end{widetext}
where $\sigma_{G(\omega_i)}$ is the uncertainty in $G(\omega_i)$, $\sigma_{V^\text{meas}}$ and $\sigma_{V^\text{meas}_0}$ are the measurement errors in the measured covariance matrices with and without pumps, respectively, and $\sigma_{V^\text{quant}_0}$ are the errors in the pump-off quantum covariance matrix.
To estimate $\sigma_{V^\text{quant}_0}$ we apply the chain rule to Eqs.~\eqref{eq:Vq-rec}~and~\eqref{eq:Vqoff}, obtaining 
\begin{widetext}
\begin{equation}
    \sigma^{V^\text{quant}}_{0\,ij} = \sqrt{\left( \dfrac{\partial V^\text{quant}_{0\,ij}}{\partial G(\omega_i)} \right)^2 \sigma_{G(\omega_i)}^2 + \left( \dfrac{\partial V^\text{quant}_{0\,ij}}{\partial \bar{n}_i} \right)^2 \sigma_{\bar{n}_i}^2 + \left( \dfrac{\partial V^\text{quant}_{0\,ij}}{\partial V^\text{meas}_{0\,ij}} \right)^2 \sigma_{V^\text{meas}_{0\,ij}}^2 + 2 \left( \dfrac{\partial V^\text{quant}_{0\,ij}}{\partial G(\omega_i)} \right) \left( \dfrac{\partial V^\text{quant}_{0\,ij}}{\partial \bar{n}_i} \right) \sigma_{G(\omega_i), \bar{n}_i}}
\end{equation}
that for $i=j$ it reads:
\begin{equation}
    \sigma^{V^\text{quant}}_{0\,ii} = \sqrt{ \left(\dfrac{G(\omega_i)^{-2} V^\text{meas}_{0\,ij}}{2\bar{n}_i + 2}\right)^2 \sigma_{G(\omega_i)}^2 + \left(\dfrac{G(\omega_i)^{-1} V^\text{meas}_{0\,ij}}{2(\bar{n}_i + 1)^2} \right)^2 \sigma_{\bar{n}_i}^2 + \left( \dfrac{G(\omega_i)^{-1}}{2\bar{n}_i + 2} \right)^2 \sigma_{V^\text{meas}_{0\,ij}}^2 - \dfrac{G(\omega_i)^{-3} \left(V^\text{meas}_{0\,ij}\right)^2}{2(\bar{n}_i + 1)^3} \sigma_{G(\omega_i), \bar{n}_i} },  \nonumber
\end{equation}
while for $i \ne j$ it becomes:
\begin{equation}
    \sigma^{V^\text{quant}}_{0\,ij} = \sqrt{ \left(G(\omega_i)^{-2} V^\text{meas}_{0\,ij}\right)^2 \sigma_{G(\omega_i)}^2 + G(\omega_i)^{-2} \sigma_{V^\text{meas}_{0\,ij}}^2}.  \nonumber
\end{equation}
For the nullifier, the uncertainty is estimated using error propagation through Eq.~\eqref{eq:DNi2_xplct}:
\begin{equation}
    \sigma_{\Delta N_i}^2 = \sigma_{V^\text{quant}_{i+1\,i+1}}^2 + \sum_{j,k} \sigma_{V^\text{quant}_{jk}}^2 + \sum_{j} \left( \sigma_{V^\text{quant}_{i+1 \, j}}^2 + \sigma_{V^\text{quant}_{j \, i+1}}^2 \right).
    \label{eq:error-nullifier}
\end{equation}
\end{widetext}

\section{Losses and Squeezing} \label{sec:NumAnalysis}

We perform a numerical analysis of two-mode squeezing as a function of pump power and losses.
The JPA Hamiltonian is expanded in the mode basis and its time evolution is computed by numerical solution of the Lindblad master equation. 
The JPA Hamiltonian is defined as~\cite{yamamoto_principles_2016}
\begin{equation}
    H = \omega_0 \left(A^\dagger A + \frac{1}{2}\right) + \frac{\omega_0}{2} g_p(t)(A^\dagger + A)^2.
\end{equation}
We define the pump signal as characterized as a sum of pure frequency tones
\begin{equation}
    g_p(t) = \sum_k A_k \cos(\Omega_k t + \phi_k) = \sum_k g_k\, e^{i\Omega_k t} + g_k^*\, e^{-i\Omega_k t},    
\end{equation}
where $g_k = \frac{A_k}2{2} e^{-i\phi_k}$ is the pump strength.
We focus on a single pump tone at frequency \(\Omega_0 = 2\omega_0\) and complex amplitude $g$.
Expanding the ladder operators on the frequency-mode basis
\begin{equation}
A(t) = \sum_i a_i\, e^{i\omega_it}, \quad A^\dagger(t) = \sum_i a_i^\dagger\, e^{-i\omega_it},    
\end{equation}
and performing a change of reference in the rotating frame of the resonator, we obtain
\begin{align}
    H &= \omega_0 \sum_{i,j} a_i^\dagger a_j \, e^{i(\omega_j - \omega_i)t} +
     \nonumber\\
    &+ \frac{\omega_0}{2}  \sum_{i,j} \left(ge^{i \Omega_0 t} + g^*e^{-i \Omega_0 t}  \right) \left(a_i^\dagger a_j^\dagger + \, a_i a_j \right).
\end{align}
Here we neglect high-frequency terms and assume $\omega_0(1 + g_p(t))\approx\omega_0$ in the weak pump limit.
We consider only two modes equally spaced around the JPA resonance frequency $\omega_0$ with frequency separation spanning the entire bandwidth of the JPA.
The Lindblad master equation describes the dynamics
\begin{equation}
    \dot{\rho} = -\frac{i}{\hbar}[H,\rho] + \sum_{j} \gamma\, \mathcal{D}[a_j]\rho,
\end{equation}
with the dissipative superoperator \(\mathcal{D}[a_j]\rho\) defined as
\begin{equation}
    \mathcal{D}[a_j]\rho = a_j\, \rho\, a_j^\dagger - \frac{1}{2}\{a_j^\dagger a_j,\, \rho\}.
\end{equation}
Here \(\gamma\) is the total loss rate assumed identical for all modes \(j = -1,1\).

\begin{figure}[t]
    \centering
    \includegraphics[width=1\linewidth]{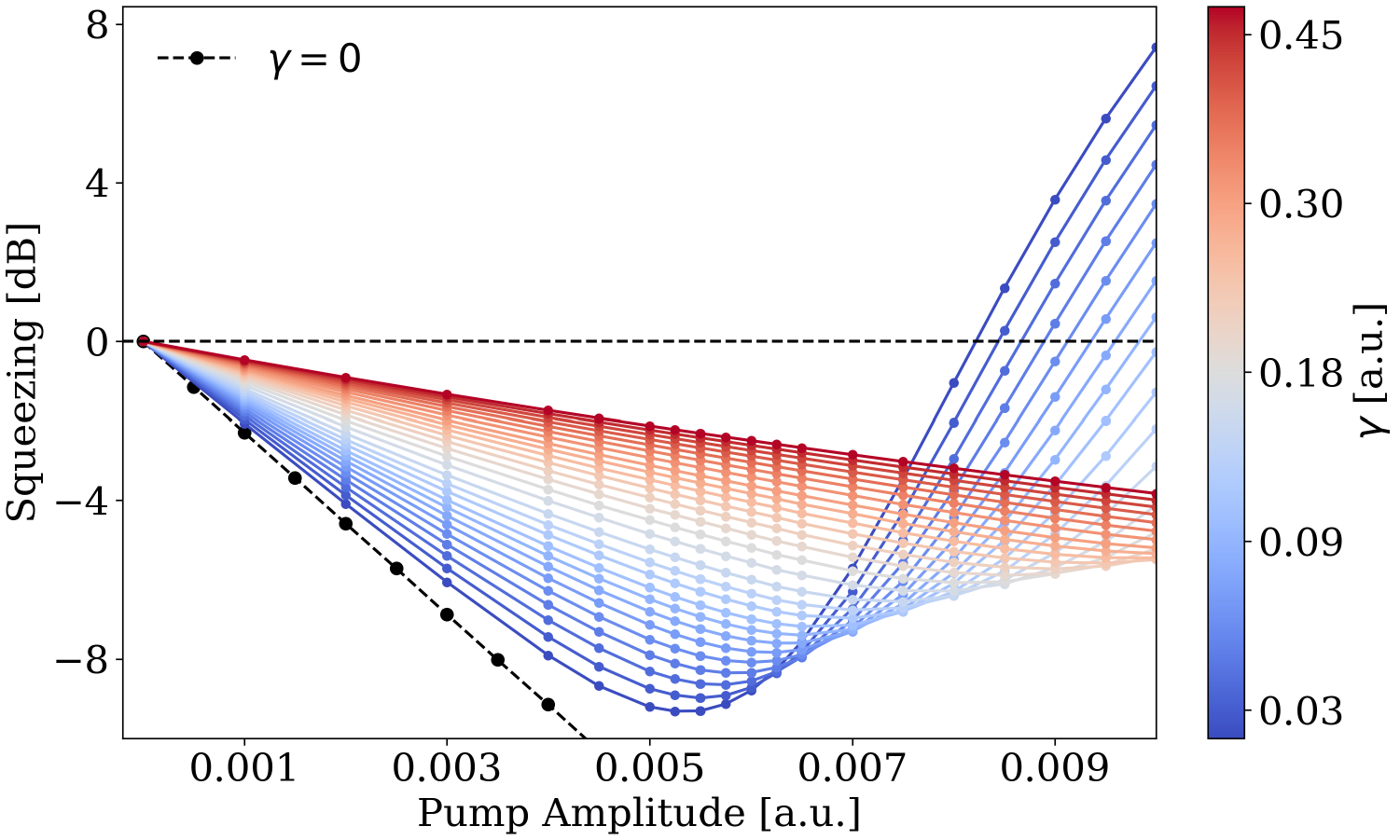}
    \caption{Squeezing as a function of the pump amplitude (in arbitrary units) for different values of the loss rate $\gamma$ indicated by the color scale. The horizontal dashed line marks the vacuum level. The black dashed-dot line refers to the lossless case $\gamma=0$.}
    \label{fig:squeezing_vs_pump_vs_gamma}
\end{figure}

This formulation consistently incorporates both the unitary and dissipative aspects of the system.
Figure~\ref{fig:squeezing_vs_pump_vs_gamma} shows the numerical computation of the squeezing as a function of pump amplitude for various values of \(\gamma\). 
The smallest eigenvalue of the covariance matrix between the quadratures pair \(\hat{p}_i\)-\(\hat{x}_{-i}\) quantifies the squeezing.
The results are evaluated at the $\theta$ of maximum squeezing and normalized to the pump-off case, as in the experiment.
We observe qualitative agreement between the numerical results and the measured two-mode squeezing.
The presence of losses explains the existence of the minimum at a particular pump amplitude, corresponding to the maximum achievable squeezing.
In contrast, the lossless case (black dashed line in FIG.~\ref{fig:squeezing_vs_pump_vs_gamma}) exhibits a monotonically increased squeezing with pump power.
Furthermore, the numerical solutions for  $\gamma\neq 0$ show degradation of squeezing above the optimal pump amplitude, as observed in the experiment.
Our numerical solutions also confirm an imbalance between the levels of squeezing and anti-squeezing induced by the losses.

\bibliography{Refs}